\documentclass{article}

\usepackage{graphicx}

\begin{document}

\title{PENETRATING  RADIATION ON THE SEA}
\author{Note by D. PACINI; {Le Radium} {VIII}, 307 {(1910)}}
\date{Translated and commented by M. De Maria and A. De Angelis
\thanks{Thanks to Dr. Stefania De Angelis from Williams Language Solutions  for help in the translation and
editing; 
and to  N.~Giglietto, S.~Stramaglia, A.~Garuccio, L.~Guerriero, E.~Menichetti, P.~Spinelli, F.~Guerra, N.~Robotti, R. Garra, L.~Cifarelli and P.~Carlson for discussions and for material about the work of Pacini.}\\INFN and University of Udine}

\maketitle

\newcommand{\BY}[1]{{#1},}
\newcommand{\IN}[4]{{#1} \textbf{#2} (#3) #4}

\noindent {\bf{Foreword -- }}
At the beginning of the twentieth century, two scientists, the
Austrian Victor Hess and the Italian Domenico Pacini,
developed independently brilliant lines of 
research,\footnote{P. Carlson and A. De Angelis, {\em Nationalism and internationalism in science: the case of the discovery of cosmic rays,} to appear in Eur. Phys. J. H,
arXiv:1012.5068 [physics.hist-ph]; {A. De Angelis,}  {\em Domenico Pacini, uncredited pioneer of the discovery of cosmic rays,} Rivista del Nuovo Cimento 33, 713-756, 2010.}
leading to the
determination of the origin of atmospheric radiation. Before their
work, the origin of the radiation  today called ``cosmic rays'' -- was
strongly debated, as many scientists thought that these particles
came from the crust of the Earth.

The approach by Hess is well known: Hess measured the rate of
discharge of an electroscope that flew aboard an atmospheric balloon.
Because the discharge rate increased as the balloon flew at higher
altitude, he concluded in 1912 that the origin could not be
terrestrial. For this discovery, Hess was awarded the Nobel Prize in
1936, and his experiment became legendary.

One year before the conclusive experiment by Hess, in 1911,  Pacini, a researcher at the National Institute of 
Meteorology  and Geodynamics and then professor at the University of Bari,
made a series of measurements to determine the  intensity of
the radiation  while the electroscope was immersed in a box in the sea
near the Naval Academy in the Bay of Livorno (the Italian Navy
supported the research) and later in the Bracciano lake. The measures are
documented in his work \emph{La radiazione penetrante alla superficie ed in
seno alle acque}\footnote{\BY{D. Pacini} {\em La radiazione penetrante alla superficie ed in seno alle acque,} 
\IN{Nuovo Cim.} {VI/3} {1912} {93} (February 1912), translated and commented by A. De Angelis, 
 {\em Penetrating radiation at the surface of and in water,} arXiv:1002.1810 [physics.hist-ph]).}. Pacini discovered (italics are in the original) that the discharge of the oscilloscope was significantly slower
than at the surface: {``The apparatus [...]  was enclosed in a copper box to be
able to immerse in
 depth.  [...] The experiments were performed [...] with the apparatus on the surface   and immersed at a depth of 3 meters. 
[...It] appears from the results of the work described in this Note: that \emph{a sizable cause of ionization exists in the atmosphere, originating from penetrating radiation, independent of the direct action of radioactive substances in the soil.}"}

Documents testify$^1$ that Pacini and Hess knew of each other's work. Pacini died in 1934, two years before the Nobel Prize was awarded for
the discovery of cosmic rays.
While Hess is remembered as the
discoverer of  cosmic rays, the simultaneous discovery by Pacini is
forgotten by most.

Pacini expressed from his very first works on penetrating radiation\footnote{See for example  \BY{D. Pacini} \IN{Rend. Acc. Lincei} {18}{1909} {123}.} the belief that the direct action of active substances in the soil was not sufficient to explain the observed properties of penetrating radiation, and developed a coherent research line to demonstrate his thesis. One year before his conclusive article reporting the measurements in 1911, Pacini made an important step by comparing the air ionization on the sea's surface and on ground;\footnote{\BY{D. Pacini}  \IN{Ann. Uff. Centr. Meteor.} {XXXII, parte I}{1910}{}}$^,$\footnote{\BY{D. Pacini} \IN{Le Radium} {VIII}{1911} {307}.} he found that the radiation is only slightly smaller at the sea's surface, and concluded that ``results seem to indicate that {\em
a substantial part of the penetrating radiation in the air, especially the one that is subject to significant fluctuations, has an origin independent of the direct action of active substances in the upper layers the Earth's crust.\footnote{In italics in the original.}''}

Here we publish a translation of the work in Ref. $5$.
Ref. 5 is, with minor modifications,  the French translation of the original article published by Pacini in Italian (Ref. 4); the translation into French was not done by Pacini, but  it comes from a late revision of the original article by Pacini himself. We thus decided to translate and publish Ref. 5, also since in the fundamental paper in which he publishes for the first time a significant effect,\footnote{ \BY{V. Hess} \IN{Phys. Zeit.} {13}{November 1912} {1084}.} Hess quotes the partial 1908-1910 results by Pacini from the French version itself (Ref. 5). 

The measurements were done on the {\em cacciatorpediniere} (destroyer) ``Fulmine'' from the Italian Navy, which was used since 1907 by the Central Bureau of Meteorology and Geodynamics; a picture of this boat taken\footnote{\BY{L. Palazzo} \IN{Bollettino della Societ\`a Aeronautica Italiana}{13}{1908}{5}.}  during the first expedition in 1907, expedition to which Pacini participated, is shown below.

%\begin{figure}
\begin{center}
\resizebox{0.67\columnwidth}{!}{\includegraphics{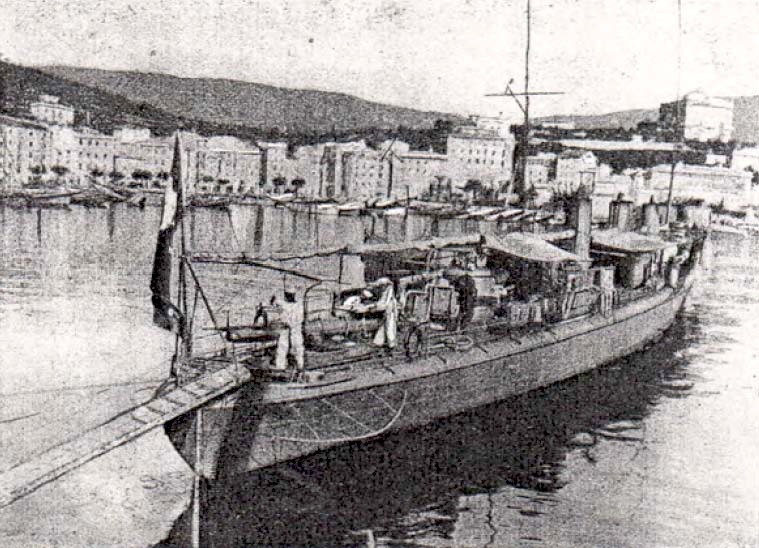} }

\end{center}

%\vskip 2mm

%\hskip 6cm Alessandro De Angelis

%\hskip 6cm Udine, , 2010

\newpage

%\hskip -8mm \includegraphics[scale=0.3]{copertina.eps}

%\newpage

\begin{center}
{\center{\bf{PENETRATING  RADIATION ON THE SEA}}}

{\center{Note by D. PACINI}}

{\center{Translated and commented by M. De Maria and A. De Angelis}}
\end{center}

\vskip 1cm

The penetrating radiation observed in the air over the soil surface is coming partly from the active substances in the upper layer of the Earth's surface as well as from their disintegration products, and partly from outside the soil.

The penetrating radiation originated from outside the soil is due at least in part to the transformation products of the radioactive elements of the air. 

The radiation originating from the soil changes from place to place with the changes of the soil nature, in proportion to the content of active materials; in the same place, it may be reduced with an increase on the permeability of the upper layers, for example because of rainfall, but, for a dry soil, it should remain 
approximately constant. 

The radiation originating from outside the soil varies with the atmospheric variations, being influenced by winds, rainfalls, modifications of the electric fields of the Earth; all these reasons can determine the accumulation of the active materials inside the lower layers of the air and the upper layers of the soil.

Many investigators, studying this phenomenon, have noticed considerable variations. For example, the author has found in Sestola in 1908 that the penetrating radiation entailed ionization rates from 6 to 30 ions per cm$^3$
 per second\footnote{D. Pacini, \textit{Rend. Lincei}, \textbf{18} (1909) 123.}. Similarly Mache,\footnote{H. Mache \textit{Sitzungsberichleu der K. Akad. der Wiss. Wien}, \textbf{119} (1910).} in his experiments conducted in Innsbruck from October 1st 1907 to October 15th 1908, reached the conclusion that the penetrating radiation due to the active substances in the air, or fallen from the air on the soil, may be 4 times greater than that coming the soil itself. 
 
Up to now, the results obtained show that if one measures the penetrating radiation in a place, with no influences from atmospheric variations or electric fields, where the soil is relatively rich in active substances, the importance of the soil and of the walls is predominant, with only minor variations in the series of values recorded. Conversely, if one performs the same measurement in a place, like Sestola, highly exposed to meteorological perturbations and variations of the electric potential, a larger variability of the penetrating radiation values is expected.  

However, according to Wulf and Kurz, most of the radiation observed in air near the ground would be due to active substances of the surface layer and the radiation that does not come from inside the ground would be negligible compared to the other. The conclusion of Wulf's work are:
\begin{enumerate}
\item The penetrating radiation is caused by primary radioactive substances lying in the uppermost layers of the Earth extending down to about 1 m below the surface.
\item The fraction of the radiation stemming from the atmosphere is so small that is impossible to be detected by the methods that have been used.
\item The time fluctuations in the $\gamma$ radiation can be explained by displacements of emanation-rich air masses under the surface of the Earth at larger or smaller depths.
\end{enumerate}

We see then that the results of experiments in Sestola, in Innsbruck and elsewhere are not in agreement with the conclusions of 
Wulf\footnote{Th. Wulf, \IN{Le Radium} {7} {1910} {171}.} and Kurz\footnote{K. Kurz,  \IN{Phys. Zeit.} {10} {1909}{834}.}. It is therefore necessary to do more research to see if the variable part of the penetrating radiation, that can sometimes take significant values, is really due to active substances contained in the soil, or whether  its origin should be searched, at least in part, outside the soil.

These facts suggested to conduct some experiments on media capable to absorb $\gamma$ radiation from the soil; for this purpose, I performed  the series of observations at the surface of the sea, which will now be discussed.

The experiments were made in Livorno, at the Naval Academy, 
with two Wulf's devices.\footnote{Th. Wulf, \IN{Phys. Zeit.} {10} {1909} {152}.} Such equipments, besides their low capacity (the ones I have used had capacities of 1.2 cm and 1.38 cm respectively) have the advantage to enable detecting possible losses caused by insulation faults.

\vskip 0.6cm
\begin{center}
{\center {\it {Comparison between the indications of the two devices}}}
\end{center}
\vskip 0.3cm

The two devices I used will be designated by the letters A and B.

I should note that the thickness of the walls of the device A is larger than  B. The loss $\Delta V_i$ due to isolation faults varied from a negligible amount, to 0.3 V per hour; only exceptionally it has reached 0.5 V. The first observations were made at the meteorological observatory of the Naval Academy. Denoting by $\Delta V$ the potential drop in volt per hour, we find that the mean of observations made with the device A $(\Delta V - \Delta V_i)$ is of 16.3 volt per hour, while the device B gives for $(\Delta V - \Delta V_i)$ the value of 25.4 V/h. These two numbers correspond, respectively, to 14.6 and 25.8 ions per cm$^3$ per second. From this value one must subtract the amount of ions which, as it will be discussed below, may be due to the direct action of the walls of the device; as a result I found:
\begin{eqnarray*}
n & = & 14.6-~4.7  = ~~9.9\mathrm{~ions~per~cm^3~and~per~second~with~device~ A}\\
n & = & 25.8 -11.0 = 14.8 \mathrm{ ~ions~per~cm^3~and~per~second~with~ device~B}\, .
\end{eqnarray*}

Some measurements with both devices were then performed in the garden of the Naval Academy.

The Academy is on the seaside. The building has on the land side a garden where I placed the two devices protected from the direct radiation from the Sun. These observations allowed me to assess how the two devices behave when placed in the same conditions. The experiments were continued for 10 days and measurements were made with each device every hour, from 7am to 8pm.

\begin{center}
\resizebox{0.7\columnwidth}{!}{\includegraphics{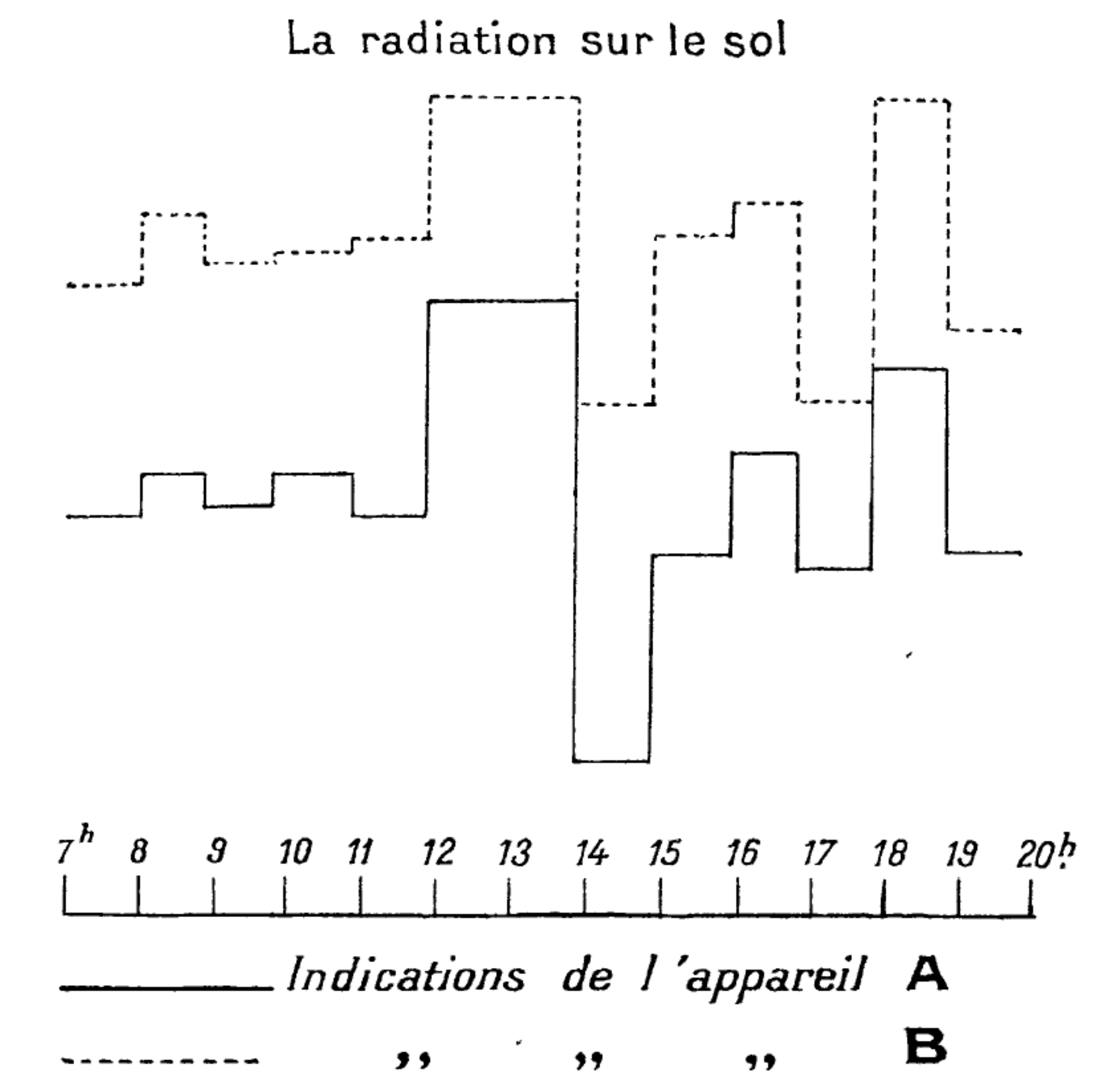} }

{Figure 1}
\end{center}

The diagram in Figure 1 displays the averages obtained taking into account all observations, and shows the time dependence of the phenomenon of penetrating radiation into this place during the 10 days in which observations were performed.
The same diagram allows us to compare the indications of the two devices operating simultaneously in the same conditions.

If we consider the complexity of the phenomenon studied, that is influenced, as we know, by the device itself, we conclude \textit{that both devices follow, with sufficient agreement, the variations of the penetrating radiation the atmosphere}.

To see how much of the observed ionization may be attributed to radiation coming from outside, it is necessary to minimize the action of external agents and to do this the two devices were placed in various locations, shielded by thick lead walls. They were exposed to the sea and on the Bracciano Lake at a distance from the shore large enough to totally exclude the action of the soil by the air layer interposed, and with a depth of the water large enough
that the action of the sea (lake) bed was negligible.
Furthermore the device A was immersed in the water of the Bracciano lake. However the absorbing layers were not successful in reducing the internal ionization below the smallest values obtained in the series of experiments on the surface of the sea.

The minimum value for $(\Delta V-\Delta V_i)$ obtained on the sea with the device A, was of $5.3$~V, which is equivalent to 4.7 ions per cm$^3$ per second. The device B measured a  minimum of $10.7$~V,  {\em{i.e.,}} 11.0 ions per cm$^3$ per second.

We can therefore say that the walls of the device A generate by themselves at most 4.7 ions per second per cm$^3$ of air contained within in device, and the walls of  device B at most 11.
The differences between the results of isolated observations and and the numbers above can be attributed to the action of penetrating $\gamma$ radiation from outside and to the action of secondary radiation generated by the first interacting with the walls of the device.

From the experiments done in the garden of the Academy, it follows that the device provides a minimum of 6.5 ions, a maximum of 16.6 and an average of 14.6 per cm$^3$ per second. Subtracting the 4.7 ions that we assume to be due to the walls, we have for the device A on the ground:

\vspace{3 mm}
\begin{tabular}{ l  l  l}
  Minimum:    & 1.8    ions per cm$^3$ and per second; & \\
  Maximum:  & 11.9 & \\
  Average:  & 6.9 &  \\
\end{tabular}
\vspace{4 mm}

\noindent and for the device B also on the ground:

\vspace{3 mm}
\begin{tabular}{ l  l l}
  Minimum:    & 3.6 ions per cm$^3$ and per second; & \\
  Maximum:  & 25.8  & \\
  Average:  & 14.1 &  \\
\end{tabular}
\vspace{3 mm}

We see that for the first unit the oscillations of the values of the penetrating radiation over the ground are such that the minimum is $26\%$ of the average and the maximum is $172\%$
of the same average.

For the second device, the minimum is $25\%$ of the average value and the maximum is $169\% $.

We can thus conclude that \textit{both devices operating simultaneously on the ground showed some oscillations in the values of the penetrating radiation that are in a great approximation of the same magnitude}.

\vskip 0.6cm
\begin{center}
{\center {\it {Observations on the sea}}}
\end{center}
\vskip 0.3cm

With one device (device B), we continued our observations in the garden of the Academy. The other (A) was put on a dinghy of the Navy and protected from direct radiation of the Sun. The dinghy was anchored at more than 300 meters from shore, a distance capable to reduce to a fraction 
of less than 4 per 100\footnote{K. Kurz, \IN{Phys. Zeit.} {10} {1909} {832}.} the radiation coming from the mainland; 
as at the point chosen for anchoring the depth of the sea was beyond 4 meters, the radiation from the seabed was completely 
absorbed\footnote{Wright, {\em {Phil. Mag.,}} February 1909.}.

This first series of observations spanned nine days between August 3 and August 19. Because of unavoidable circumstances, bad weather
in particular, we had to suspend the tests for a few days.

The result of this series of observations was that the device A measured a minimum of 4.7 ions, a maximum of 15.2 and an average of 8.9.

Since the minimum of 4.7 is compatible with the radiation from the walls, we find that:

%\begin{fleqn}
%Text and stuff.
%
%\begin{align}
%Equation aligned left.
%\end{align}
%
%Bla bla.
%
%\end{fleqn}
%

\[
a)\mathrm{~on~the~sea~(device~A)}\left\{\begin{array}{l c c r}
	\mathrm{minimum ~very~low} & ~~~~& ~~~~~\\
	\mathrm{maximum ~of~10.5~ions} & ~~~~& ~~~~~\\
	\mathrm{average~of~4.2}& ~~~~& ~~~~~
	   \end{array}\right.
\]
\noindent while
\[
\mathrm{~on~the~ground~(device~A)}\left\{\begin{array}{l c c r}
	\mathrm{minimum ~1.8~ ions} & ~~~~& ~~~~~\\
	\mathrm{maximum ~of~11.9~ions} & ~~~~& ~~~~~\\
	\mathrm{average~of~6.6~ions}& ~~~~& ~~~~~
	   \end{array}\right.
\]

\noindent where we took into account all the observations made on land with the device A. 
In average the unit A measures  on the sea 2.4 ions less than on the land.

We can compare the the oscillations observed on the sea and those found on ground during the same period with the device B. 
Device B measured in the period from 3 August 19: 

\[
a')\mathrm{~device~B~at~ground~}\left\{\begin{array}{l c c r}
	\mathrm{minimum ~7.8~ ions} & ~~~~& ~~~~~\\
	\mathrm{maximum ~of~19.9~ions} & ~~~~& ~~~~~\\
	\mathrm{average~of~12.8~ions}& ~~~~& ~~~~~
	   \end{array}\right.
\]

Considering together the data $a$ and $a'$ we see that the maximum value obtained on the sea is $ 255\%$ of the average, and the maximum value obtained simultaneously on land is $155\%$
of the average. 

Conclusion: while both devices showed on the ground the same oscillations of the penetrating radiation, now the device A shows oscillations that are clearly larger than those measured in the same time on the ground using the device B. 

From August 20 to August 26 the position of the devices was exchanged: B was transferred on the sea and A in the garden of the Academy. 
The device B on the sea measured a minimum of 11.0 ions, a maximum of 20 and an average of 19.4, and assuming that 11 ions are due to the device itself, we find that: 

\[
b)\mathrm{~on~the~sea~with~the~device~B~}\left\{\begin{array}{l c c r}
	\mathrm{minimum ~very~low} & ~~~~& ~~~~~\\
	\mathrm{maximum ~of~15~ions} & ~~~~& ~~~~~\\
	\mathrm{average~of~8.4}& ~~~~& ~~~~~
	   \end{array}\right.
\]
\noindent while
\[
a)\mathrm{~on~the~ground~with~the~device~B}\left\{\begin{array}{l c c r}
	\mathrm{minimum ~of~3.6} & ~~~~& ~~~~~\\
	\mathrm{maximum ~of~23.8~ions} & ~~~~& ~~~~~\\
	\mathrm{average~of~13.4~ions}& ~~~~& ~~~~~
	   \end{array}\right.
\]

\noindent taking into account again of all the observations made in the land with B. 

In average the unit B measures on the land 5 ions more than on the sea.

For data on land from August 20 to August 26: 

\[
b')\mathrm{~on~the~ground}\left\{\begin{array}{l c c r}
	\mathrm{minimum 3.0} & ~~~~& ~~~~~\\
	\mathrm{maximum ~of~10.0~ions} & ~~~~& ~~~~~\\
	\mathrm{average~of~6.4}& ~~~~& ~~~~~
	   \end{array}\right.	   
\]

Considering all the data $b$ and $b'$ we see that the maximum obtained at sea is $178\%$
of the average value, the maximum obtained in the same time on the land being $172\%$
of the corresponding average. This second series of experiments indicates that the penetrating radiation on the sea undergoes oscillations that are at least of the same order of magnitude as on the land.

About \textit{the evolution of the phenomenon on the sea surface and on the land,} I notice that, when the two devices operate on the land under the same conditions, Figure 1 reveals for both the same trend of the penetrating radiation during the ten days of observation, and it would be of great interest to compare the trend on land and sea; it would be sufficient to compare the series of observations made on the sea with the series on the land during the same period. But it is clear that in order to show the existence of a possible correlation, it is necessary to do a long series of measurements in order to reduce the errors due to unavoidable local influences, determined by many occasional causes, to a minimum. The winds that can transport variable amounts of active material in the vicinity of the devices, the location of the instruments, the holders and the shelters can be sources of error. On the sea, at a suitable distance from the shore, these sources of error can be reduced; in any case,  to make a significant comparison, a period of time longer than that I dedicated to the experiment would be needed.

However, I plotted the results of the observations made on the sea with the device A, and on the land with the device B, for the first and longest series of measurements made simultaneously at the sea surface and on land of from 3 to 19 August. The diagram in Figure 2 was constructed taking in account the mean values at every hour. Apart from the very high value obtained between 7:00 and 8:00 am on the land with device B, a correlation between the two trends can be perceived. But I repeat that for deciding on this question, a longer series of observations would be necessary.

\begin{center}
\resizebox{0.7\columnwidth}{!}{\includegraphics{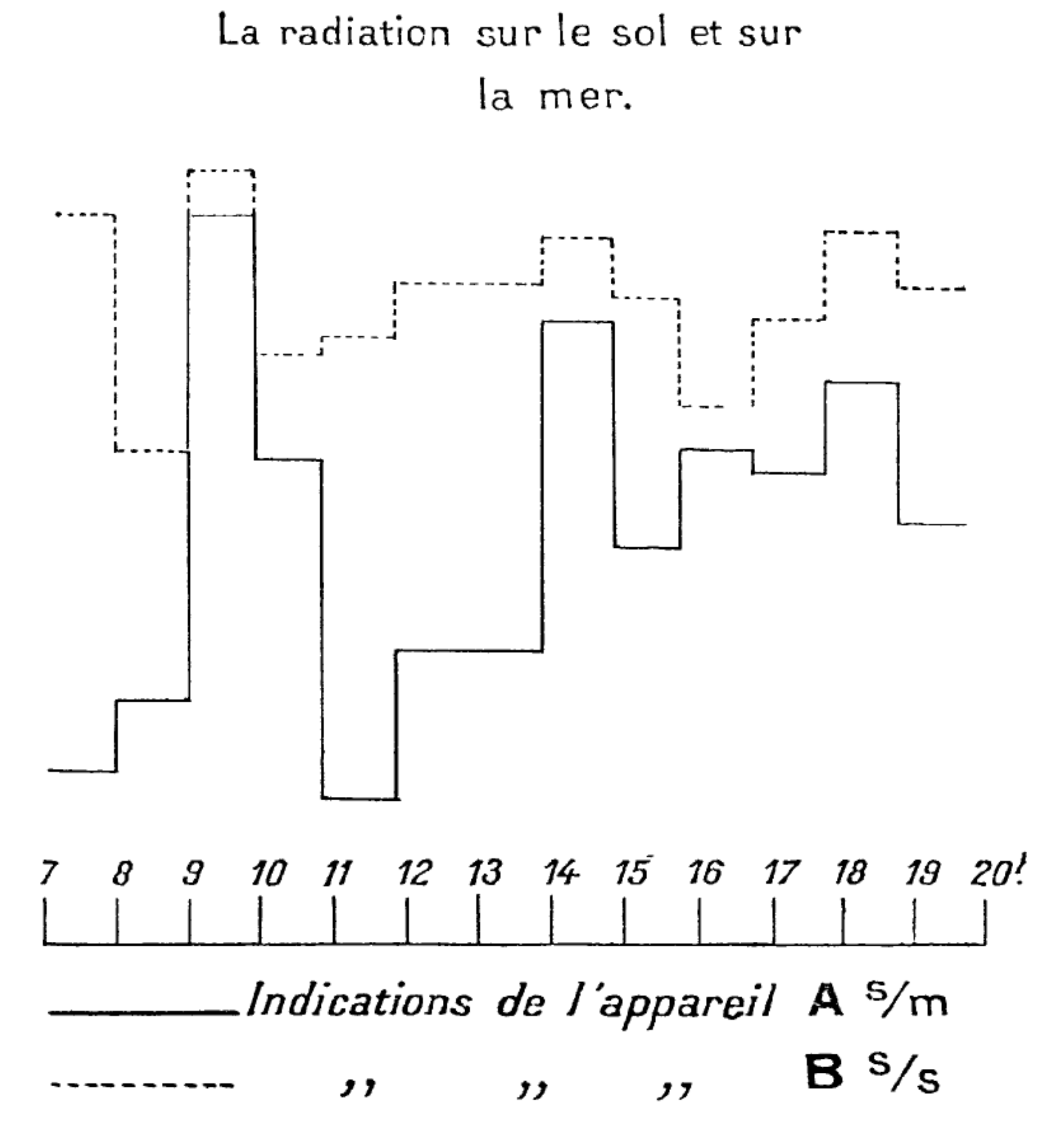} }

{Figure 2}
\end{center}

We now want to suggest a possible explanation of the noticeable oscillations observed at the sea surface. The radiation observed on the sea, 
as far as we know, can not be attributed to the active materials in the air and to the products of disintegration of the radioactive emanation contained in the sea, products whose action on the surface could, as noted by Kurz, be spread in a more effective way due to the effect of the sea waves. We found however from our observations, that there is no relationship between the state of the sea and the value of the penetrating radiation, so \textit{we do not have any evidence to attribute oscillations of this magnitude to the decay products of radioactive emissions released into the sea by the effect of waves.} 

About this result, I should recall another result that I have obtained in the Gulf of Genova\footnote{D. Pacini,  \IN{Nuovo Cim.}{15}{1908}{18}.}, whre, studying the ionization on the open sea, I found that often such ionization was relatively low in rough sea conditions.

During the days in which the experiment was performed, the prevailing winds blew from the west, that is to say, from the sea; there were anyway also observations made under winds from the coast.

An examination of the results shows that with low or very low winds from the coast, the values of the penetrating radiation tend to be below the average, and that with moderate winds from the sea there is also a predominance of values below the average.

Note that no observations were made in presence of strong winds. Moreover, when observations were suspended because of atmospheric disturbances,  violent sea winds were often blowing.
Sometimes the observations began few hours or even few moments after the wind had stopped, and one can assume in such conditions that the physical properties of air, at least at the sea surface, must be influenced for a certain amount of time by the conditions determined by the winds that previously had blown so violently. A long series of measures would be required to determine if on the sea, at a certain distance from shore, a difference of action related to the difference of the wind direction and speed exists.

%\newpage

\vskip 0.6cm
\begin{center}
{\center {\it {Summary}}}
\end{center}
\vskip 0.3cm

\begin{enumerate}
\item The number of ions due to penetrating radiation on the sea is estimated to be 2/3 of that on the ground, and  consistent values for this ratio have been measured using two different devices.
\item Under the experimental conditions described in this work and according to the measurements provided by the device B, whose walls have a thickness suitable to allow the passage of the largest part of highly penetrating radiation,  the number of ions generated  per cm$^3 $
per second by this radiation at the sea surface, is smaller by 5 ions with respect to the average value observed on land. We assumed for the charge of the ions the new value $e = 4.65 \times 10^{-10}$ U.E.S.
\item The penetrating radiation on the sea at a distance of more than 500 meters from shore with water depths larger than 4 meters, under conditions that allow to neglect the radiation from the soil, undergoes oscillations that are at least of the same order of magnitude than observed at the same time on the ground.
\item As no relationship appears to exist between the state of the sea and the action of gamma rays, we cannot attribute the oscillations of the radiation level to decay products of emanations that could possibly be released from the water with larger intensity, according to the intensity of the sea waves related to storms.
\item During the experiments, the winds blowing were generally weak and came from the sea; the data collected do not allow establishing whether a
relationship exists between the penetrating radiation  and the wind's speed and direction.
\end{enumerate}
These results agree with those  obtained by the author in Sestola and later by  Mache in Innsbruck: in the air at the sea surface, as well as  on the rocks of the Apennines near Modena,
I found that the penetrating radiation may undergo large oscillations from relatively high values to so small values that they could be attributed almost entirely to the action of the walls of the device. Large oscillations occur near the rocks of the mountains, on the air above the sea, and on the ground, where the ionization produced by radiation is larger because of the direct action of active substances from the soil.

The number of ions due to penetrating radiation on the sea being the 2/3 of those on land, we can explain the relatively high value of the ionization of the free atmosphere at the surface 
of the sea\footnote{D. Pacini, \textit{Nuovo Cim.,} \textbf{15} (1908) 18.}. An important question remains open, since our knowledge of the quantity of active substances in sea water and air does not allow us to explain the large values found for the penetrating radiation on the sea,\footnote{A.S. Eve, \textit{Terr. Magn. and Atm. Elect.,} 1910 -- D. Pacini, \textit{Nuovo Cimento}, (1910) 449.} nor on the mainland (Gockel's balloon 
measurements\footnote{A. Goeckel, \textit{Phis. Zeit.,} (1910) 280.} and Wulf's measurements on the Eiffel Tower\footnote{A. Wulf, \textit{Phis. Zeit.,} (1910) 814.}) at a sufficient height that one can neglect the action of the active substances form the soil. Anyway such results seem to indicate that {\em
a substantial part of the penetrating radiation in the air, especially the one that is subject to significant oscillations, has an origin independent of the direct action of active substances in the upper layers the Earth's crust.}

%[Manuscript received April 2, 1911. Translated by L. Bloch]

\vskip 5mm

 $\mathrm{\left[Manuscript\: received\: April\: 2,\: 1911\right].}$
 
 \hskip 5cm $\mathrm{\left[Translated\: by \:L.\: Bloch\right].}$

\vspace{2cm}

\hspace{-0.5cm}\begin{center}
\resizebox{1.1\columnwidth}{!}{\includegraphics{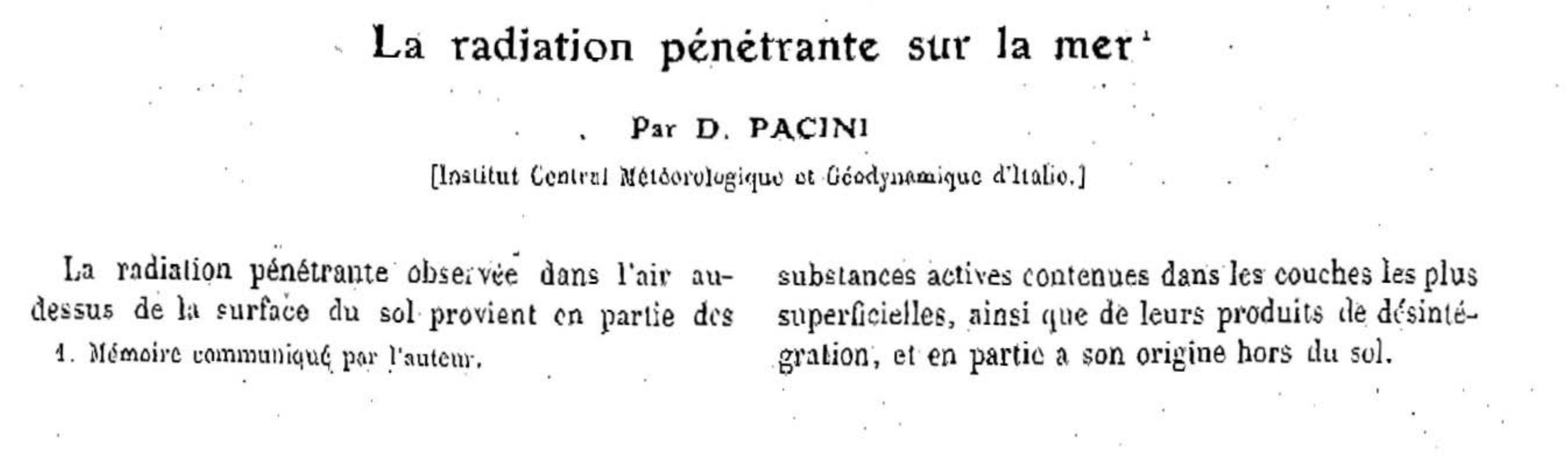} }
\end{center}

\pagestyle{empty}
\nonumber

\vspace*{-4cm}\hspace*{-4.5cm} \includegraphics{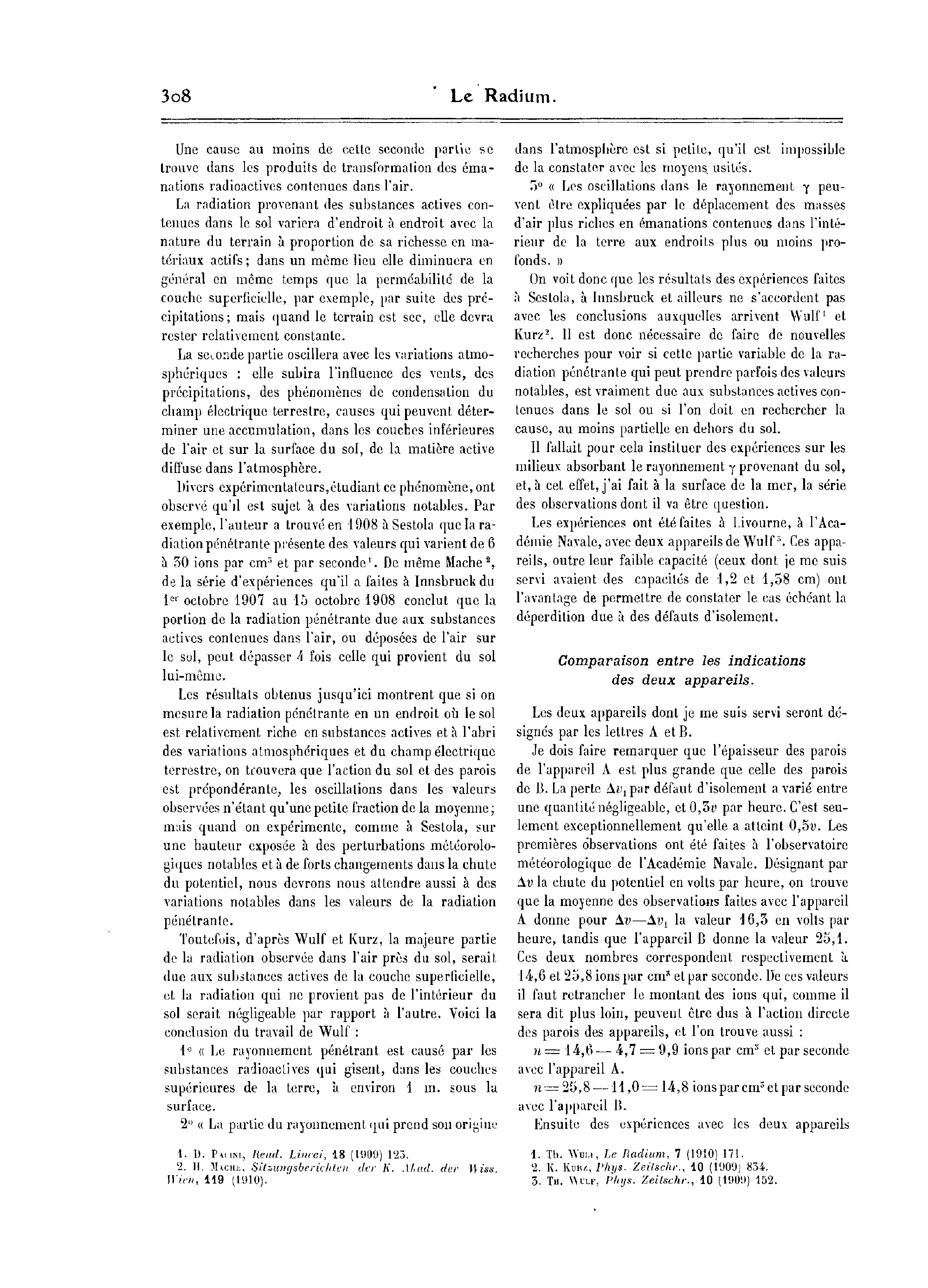} \newpage
\vspace*{-4cm}\hspace*{-3.5cm}\includegraphics{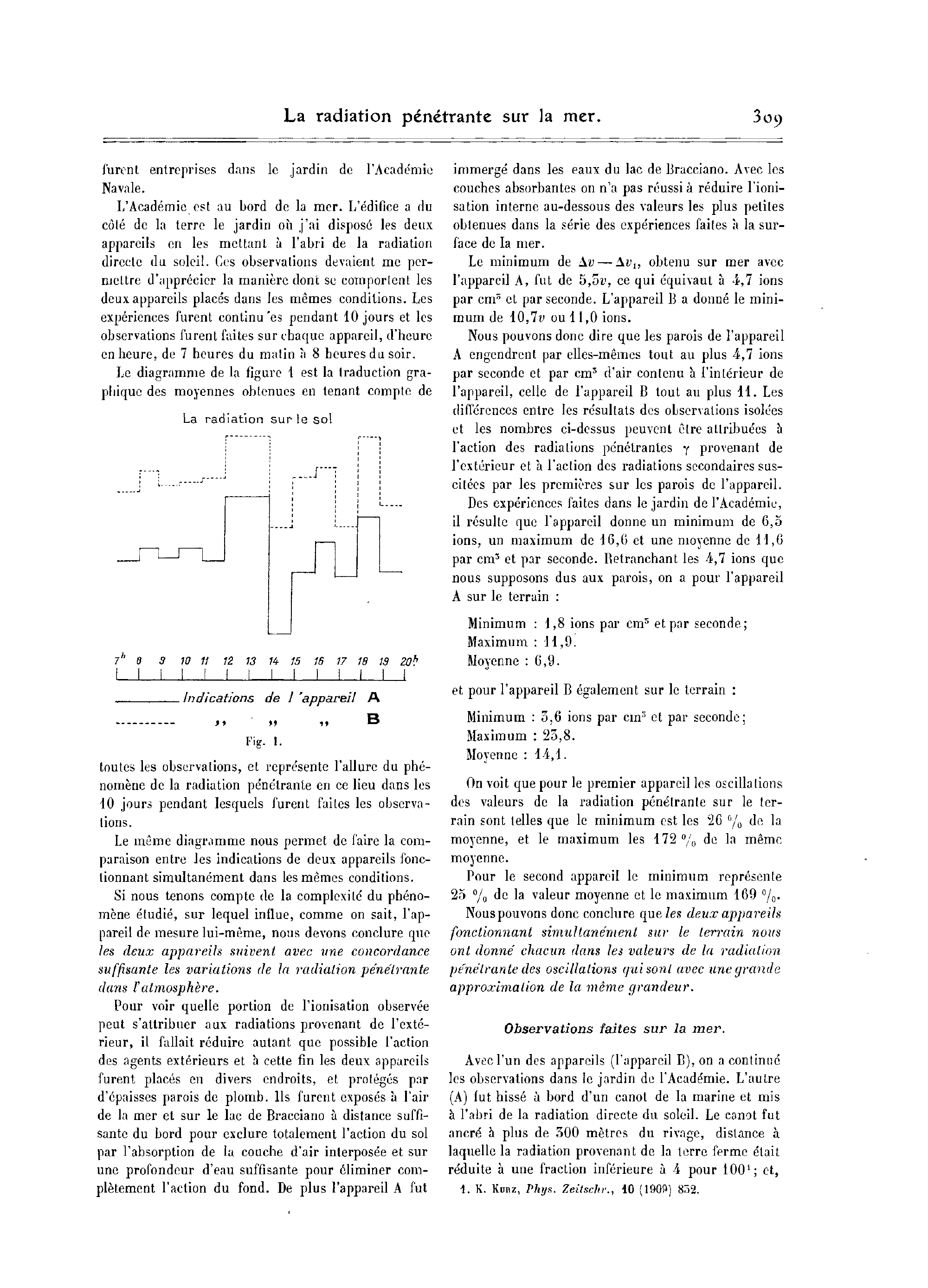} \newpage
\vspace*{-4cm}\hspace*{-4.5cm}\includegraphics{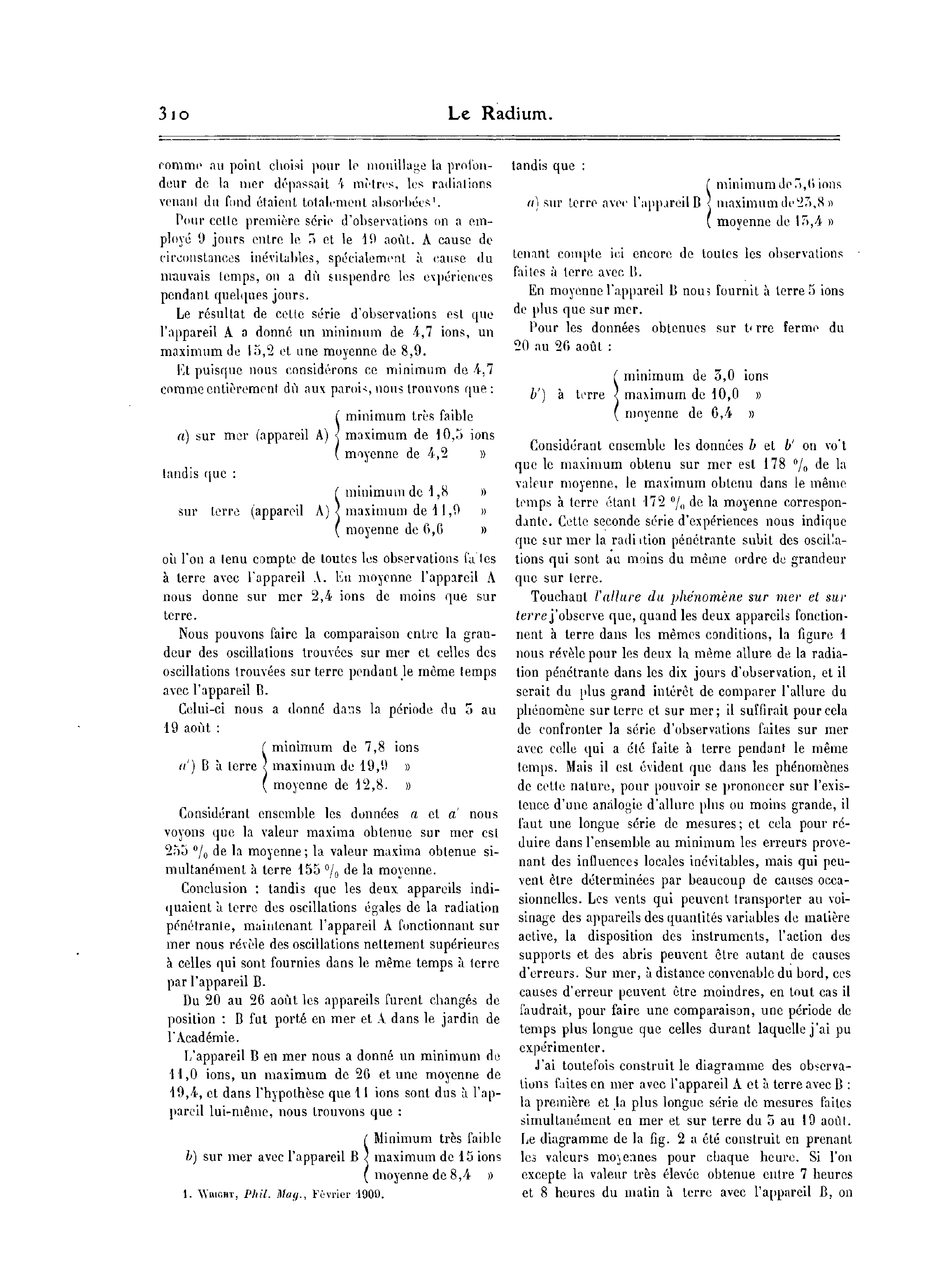} \newpage
\vspace*{-4cm}\hspace*{-3.5cm} \includegraphics{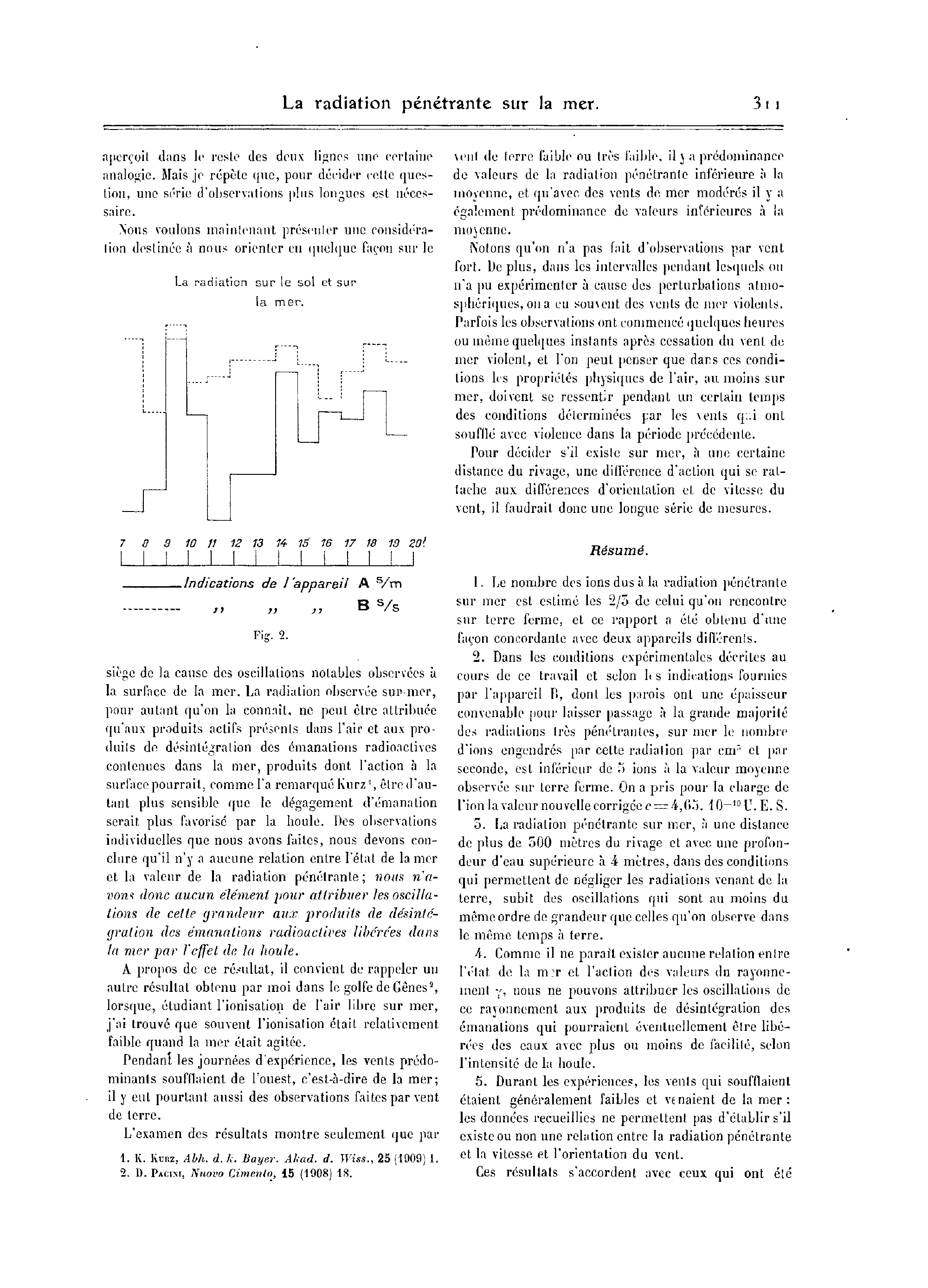} \newpage
\vspace*{-4cm}\hspace*{-4.5cm} \includegraphics{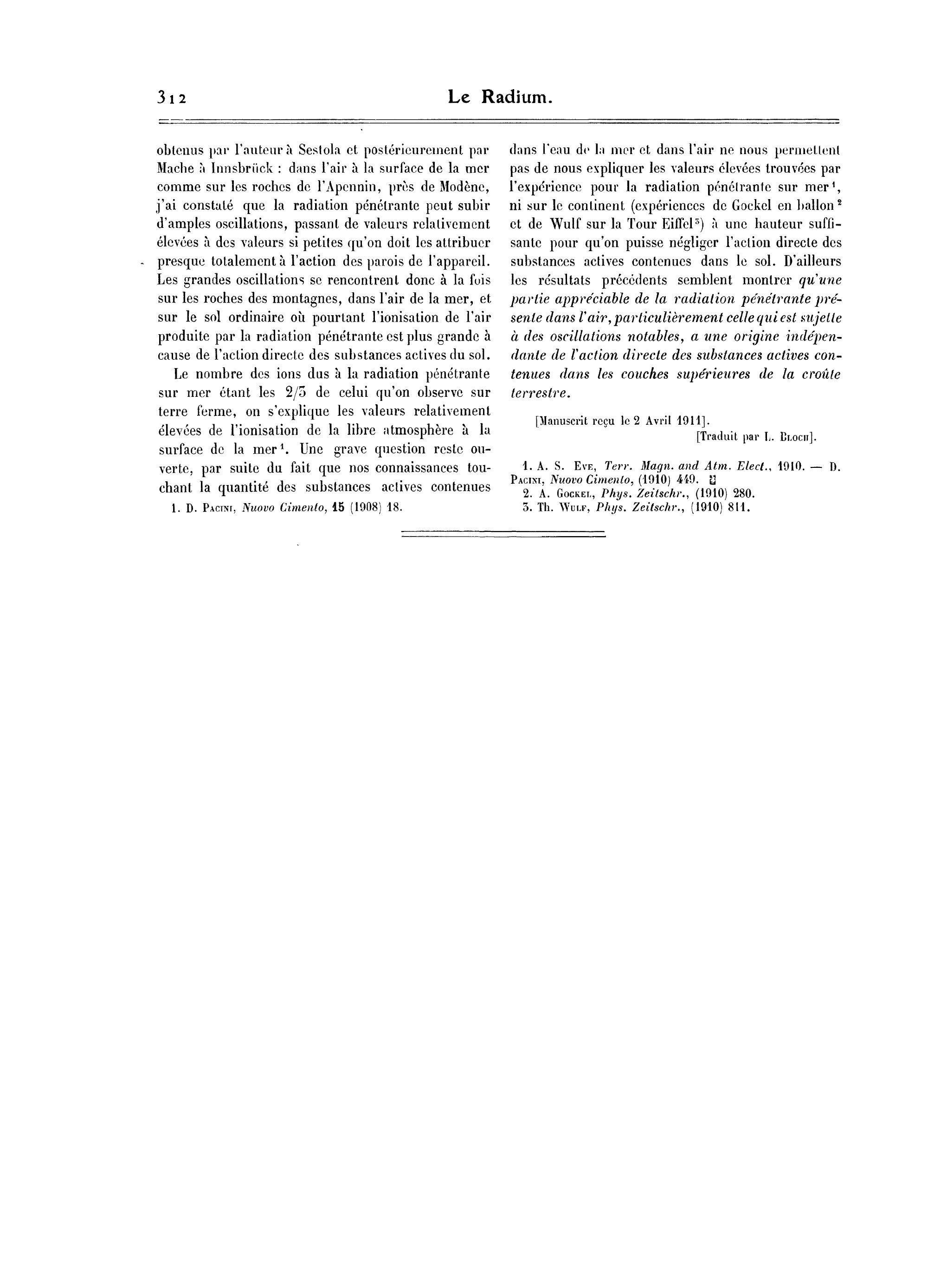} \newpage
%\hspace{-2.1cm}\includegraphics{NC12_3_62.pdf} \newpage
%\hspace{-2.1cm}\includegraphics{NC12_3_63.pdf}\newpage
%\hspace{-2.1cm}\includegraphics{NC12_3_64.pdf}\newpage
%\hspace{-2.1cm}\includegraphics{NC12_3_65.pdf}\newpage
%\hspace{-2.1cm}\includegraphics{NC12_3_66.pdf}\newpage
%\hspace{-2.1cm}\includegraphics{NC12_3_67.pdf}\newpage
%\hspace{-2.1cm}\includegraphics{NC12_3_68.pdf} 

\end{document}